\documentclass{article}

\PassOptionsToPackage{numbers,compress}{natbib}
\usepackage[preprint]{neurips_2025}

\usepackage[utf8]{inputenc}
\usepackage[T1]{fontenc}
\usepackage{amsmath,amsthm,amssymb,amsfonts}
\usepackage{graphicx}
\usepackage{booktabs}
\usepackage{algorithm}
\usepackage{algorithmic}
\usepackage{hyperref}
\usepackage{url}
\usepackage{subcaption}
\usepackage{xcolor}
\usepackage{nicefrac}
\usepackage{microtype}

\title{CodeEvolve: LLM-Driven Evolutionary Optimization with Runtime-Enriched Target Selection for Multi-Language Code Enhancement}

\author{
\begin{tabular}{c}
Ajay Krishna Borra$^{*}$ \quad
Wenzhuo Yang$^{*}$ \quad
Samarth Arora$^{*}$ \quad
Akhilesh Deepak Gotmare \\
Gokulakrishnan Gopalakrishnan \quad
Tharun Gali \quad
Madhav Rathi \quad
Doyen Sahoo \\
Manpreet Singh \quad
Mayuresh Verma \quad
Laksh Venka \quad
Shuchita Singh \\
Salesforce \\
{\small \texttt{wenzhuo.yang@salesforce.com} \quad
\texttt{samartharora@salesforce.com}} \\
{\small \texttt{akhilesh.gotmare@salesforce.com} \quad
\texttt{ggopalakrishnan@salesforce.com}} \\
{\small \texttt{tgali@salesforce.com} \quad
\texttt{madhav.rathi@salesforce.com} \quad
\texttt{dsahoo@salesforce.com}} \\
{\small \texttt{manpreet.singh@salesforce.com} \quad
\texttt{mayuresh.verma@salesforce.com}} \\
{\small \texttt{lvenka@salesforce.com} \quad
\texttt{shuchita.singh@salesforce.com}} \\
$^{*}$Primary authors
\end{tabular}
}

\begin{document}

\maketitle

\begin{abstract}
We present CodeEvolve, an evolutionary framework for improving program performance and code quality with Large Language Models (LLMs). CodeEvolve extends OpenEvolve~\cite{openevolve} with runtime-guided target selection, Monte Carlo Tree Search (MCTS), automated code refinement, and language-specific evaluation pipelines for Java and Salesforce Apex. The system uses Java Flight Recorder (JFR) profiles to build weighted component graphs and select optimization targets that account for most execution cost, reducing reliance on manual bottleneck identification. For each target, CodeEvolve generates candidate edits, evaluates them through build validation, unit tests, performance checks, static analysis, and LLM-based review, and retains only variants that preserve functional correctness. Across real-world optimization tasks, CodeEvolve improves performance and code metrics while maintaining correctness. On a large enterprise Java codebase, it achieves an average speedup of 15.22$\times$ across seven hotspot functions and outperforms single-pass LLM optimization on five of them. An ablation study on Apex optimization shows that the full MCTS-augmented configuration produces 19.5 valid programs out of 20 on average, indicating that search, filtering, and refinement each contribute to more reliable optimization.
\end{abstract}

\section{Introduction}

Optimizing production code is difficult because the best change is rarely determined by performance alone. Engineers must also preserve behavior, respect language and platform constraints, and keep the resulting code maintainable. Static analysis, rule-based transformations, and manual review remain useful, but they do not scale easily to large, heterogeneous codebases where performance bottlenecks are unevenly distributed and correctness regressions are costly.

Large Language Models (LLMs) make it possible to automate parts of this workflow. They can interpret code, propose syntactically valid changes, and reason about algorithmic alternatives. Used naively, however, they also introduce new failure modes: generated edits may not compile, may overfit to a narrow prompt, or may improve a local metric while breaking behavior elsewhere. A practical optimization system therefore needs more than a model call. It needs a way to choose the right targets, search across multiple candidate edits, evaluate them objectively, and reject changes that do not preserve correctness.

Target selection is especially important. Real applications often have skewed execution profiles, with a small number of paths accounting for most runtime~\cite{ball1994optimally,knuth1971empirical}. Optimizing cold code consumes model budget and engineering attention without much benefit, while optimizing hot paths can produce visible performance gains. CodeEvolve addresses this problem by coupling runtime profiling with evolutionary search, so that the optimizer focuses on code regions with measured execution impact.

\subsection{Motivation and Challenges}

Contemporary software development environments present several key challenges for automated code optimization:

\textbf{Scale and Complexity:} Enterprise-scale codebases routinely comprise millions of lines of code across multiple programming languages and frameworks, rendering manual optimization infeasible. Automated solutions must therefore accommodate heterogeneous language features, architectural styles, and optimization objectives at scale.

\textbf{Multi-Objective Optimization:} Meaningful code optimization inherently involves balancing competing objectives, including execution performance, memory efficiency, code readability, maintainability, and test robustness. Existing optimization techniques typically emphasize isolated metrics, thereby overlooking critical trade-offs among these dimensions.

\textbf{Language-Specific Constraints:} Programming languages impose distinct constraints and optimization considerations. For example, Salesforce Apex development must adhere to strict governance limits and platform-specific execution models, whereas Java optimization requires careful consideration of JVM behavior, memory management, and enterprise architectural patterns.

\textbf{Correctness Preservation:} Automated optimization must preserve functional correctness while improving non-functional attributes. Achieving this goal requires validation mechanisms such as compilation checks, unit testing, static analysis, and semantic validation to be integrated into the optimization process.

\subsection{Our Contributions}

CodeEvolve makes the following contributions:

\begin{enumerate}
    \item \textbf{Runtime-guided target selection:} CodeEvolve uses JFR profiles to construct weighted component graphs and select performance-critical methods before optimization begins. This step directs the search toward code regions with measured execution impact.

    \item \textbf{Multi-language evolutionary optimization:} The framework supports both Salesforce Apex and Java through language-specific evaluators, configurable search parameters, and optimization objectives that can combine correctness, performance, maintainability, and code quality.

    \item \textbf{Search and refinement for reliable LLM edits:} CodeEvolve combines evolutionary search, MCTS-based exploration, and code-refinement agents to generate, repair, and retain only candidates that pass the evaluation pipeline.

    \item \textbf{Empirical evaluation on Apex and Java:} We evaluate CodeEvolve with an Apex ablation study and real-world Java hotspot optimization. The full system produces 19.5 valid Apex programs out of 20 on average and reaches 15.22$\times$ average speedup across seven Java hotspot functions.
\end{enumerate}

A key design principle of CodeEvolve is \textbf{validation-first refactoring}. Each candidate optimization is evaluated before it can influence the search population. This design limits error propagation, avoids expensive combinatorial search over unverified edit combinations, and gives the optimizer a reliable signal for selecting future candidates.

\section{Related Work}

\subsection{Evolutionary Code Optimization}

Evolutionary techniques have long been applied to program synthesis, repair, and optimization~\cite{koza1992genetic}. These methods search over program variants, but applying them to production systems is difficult because the search space is large and each candidate must be evaluated for correctness. AlphaEvolve~\cite{alphaevolve} revisits this setting with LLM-guided evolutionary programming. CodeEvolve builds on this direction by adding runtime-guided target selection, multi-language evaluation, and refinement mechanisms designed for enterprise Java and Apex codebases.

\subsection{LLM-Based Code Generation and Search}

Large Language Models such as GPT-family systems~\cite{gpt4}, Claude, and code-oriented models~\cite{brown2020language,nijkamp2023codegen} can generate and transform code from natural-language instructions. Much of this work uses single-pass or prompt-based generation, which leaves the model responsible for finding a correct and efficient edit in one attempt. CodeEvolve instead treats the LLM as a proposal mechanism inside an iterative search loop.

MCTS~\cite{Browne_MCTS_Survey} provides one way to structure this loop by balancing exploration and exploitation across candidate programs. In CodeEvolve, MCTS is used to explore code modification paths and to guide refinement of invalid candidates before they can affect the optimization population.

\subsection{Execution-Guided Refinement}

Execution feedback is commonly used to repair generated code. CodeT~\cite{chen2023codet} uses test feedback, Reflexion~\cite{shinn2023reflexion} uses verbal self-reflection, and LDB~\cite{zhang2024ldb} adds fine-grained runtime traces to support debugging on benchmarks such as HumanEval~\cite{chen2021evaluating} and MBPP~\cite{austin2021program}. RethinkMCTS~\cite{zhou2023rethinkmcts} combines MCTS with block-level feedback before code generation. These systems focus primarily on functional correctness; CodeEvolve uses similar feedback loops but optimizes for correctness and non-functional objectives, including runtime performance.

\subsection{LLM-Driven Performance Optimization}

Recent systems also optimize beyond correctness. PERFCODEGEN~\cite{li2023alphacode} improves correct LLM-generated programs by feeding runtime bottleneck information back to the model. LLaMEA~\cite{mei2024llamea} uses LLMs to generate and evolve black-box metaheuristic algorithms, echoing earlier work on evolving algorithms with Linear Genetic Programming~\cite{brameier2007linear}. CodeEvolve differs by targeting existing enterprise code, preserving behavior through cascaded validation, and using runtime profiles to choose which code should enter the optimization loop.

\subsection{Runtime Profiling for Target Selection}

Profiling has long been used to identify performance-critical code regions~\cite{ball1994optimally,knuth1971empirical}. Modern tools such as Java Flight Recorder (JFR)~\cite{oracle2023jfr} provide low-overhead method timing, call frequency, allocation, and CPU data for production-like workloads. Most LLM-driven optimization frameworks still assume that a human has already chosen the target code. CodeEvolve removes this assumption by converting JFR profiles into weighted component graphs and feeding selected hot paths into the optimization pipeline.

\section{Methodology}

\subsection{Runtime Profiling Enrichment}

CodeEvolve begins by identifying code that is worth optimizing. The profiling enrichment module combines static call relationships with JFR runtime measurements, then passes a small set of high-impact targets to the optimizer. Full construction and pruning pseudocode is provided in Appendix~\ref{app:algorithms}.

\subsubsection{Static Code Representation}

We define the \emph{Code Component Graph} (CCG) as an unweighted directed graph $G = (V, E)$ where $V$ is the set of code components (functions, methods) in the target codebase and $E \subseteq V \times V$ is the set of dependency edges such that $(u, v) \in E$ if and only if component $u$ calls or references component $v$. This function-level resolution is fine enough to isolate individual optimization targets while keeping context manageable, since each node corresponds to a self-contained compilation unit whose callers and callees can be precisely enumerated through static analysis.

\subsubsection{Runtime Profiling and Weighted Graph Construction}

The profiling enrichment module uses Java Flight Recorder (JFR)~\cite{oracle2023jfr} data from production or staging workloads. For each component, it records cumulative execution time and call frequency, along with optional allocation and CPU measurements when available.

Using this profiling data, each node $v \in V$ is annotated with a weight vector:
\[
W_v = \langle T(v),\; C(v) \rangle
\]
where $T(v)$ is cumulative execution time and $C(v)$ is total call frequency across the profiling window. This transforms the static CCG into a \emph{Weighted Component Graph} (WCG), $G_w = (V, E, W)$, where $W = \{W_v \mid v \in V\}$.

\subsubsection{Context Pruning and Target Selection}

Target selection identifies high-value optimization candidates from the WCG. A component $v$ is included in the target set $V_{\text{target}}$ if it satisfies:
\[
v \in V_{\text{target}} \quad \text{if} \quad T(v) \geq \tau_{\text{time}} \;\;\text{or}\;\; C(v) \geq \tau_{\text{freq}}
\]
where $\tau_{\text{time}}$ and $\tau_{\text{freq}}$ are configurable thresholds for execution time and call frequency, respectively.

For each target $t \in V_{\text{target}}$, the optimizer receives two kinds of context:
\begin{itemize}
    \item \textbf{Active (writable) context:} The target component $t$ itself, which the evolutionary optimizer is permitted to modify.
    \item \textbf{Frozen (read-only) context:} All immediate neighbors $n$ of $t$ in the WCG, i.e., components where $(t, n) \in E$ or $(n, t) \in E$. These provide necessary dependency information but are not modified during optimization.
\end{itemize}
All other nodes are pruned from the prompt. In practice, the frozen context includes imports, related methods sharing state or data structures, type definitions, interfaces, and constants referenced by the target. This keeps the prompt small while preserving enough dependency information for correctness.

The output of the runtime profiling enrichment module is a set of prioritized target components, each paired with its pruned context subgraph. This output is passed directly to the framework's context builder, which incorporates the information into the evolutionary optimization pipeline described in the following sections.

\subsection{Framework Architecture}

\begin{figure}
    \centering
    \includegraphics[width=\linewidth]{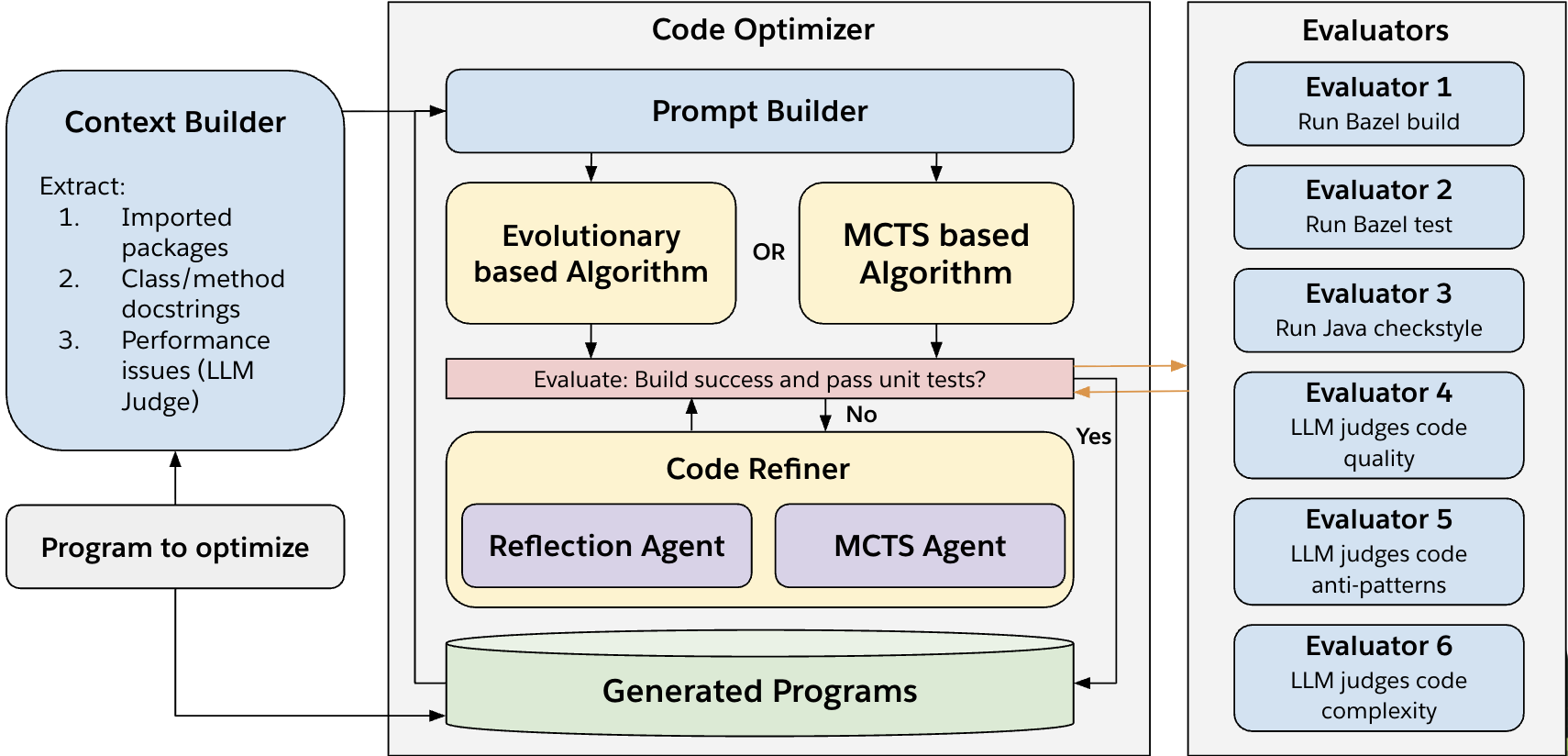}
    \caption{CodeEvolve system architecture for optimizing Salesforce Monolith Java programs.}
    \vspace{-0.1cm}
    \label{fig:1}
\end{figure}

CodeEvolve uses a modular architecture in which target selection, prompt construction, search, refinement, and evaluation are separated. Figure~\ref{fig:1} illustrates the Java configuration used for Salesforce Monolith optimization; other languages use the same control flow with different evaluators.

The core components are summarized below.

\textbf{Context Builder:}
The context builder receives prioritized targets and pruned dependency context from Section~3.1. It extracts referenced imports, signatures, constants, and profiling annotations, then packages them into the prompt so the LLM sees the writable code, nearby read-only context, and measured performance signal.

\textbf{Evolutionary Optimizer:}
The optimizer extends AlphaEvolve with context-aware prompts, population-based search, and an evolutionary-history summary that records prior edits and scores. Prompts include the current program, profiling metrics, top-performing variants, diagnostic feedback, and language-specific constraints. Appendix~\ref{app:prompt-template} shows a representative prompt structure.

\textbf{MCTS Optimizer:}
CodeEvolve also supports MCTS as a cost-aware search strategy. Each tree node represents a candidate optimized program, and node selection uses the Upper Confidence Bound (UCB) score:
\begin{equation}
\text{UCB}(n) = \frac{V(n)}{N(n)} + c \sqrt{\frac{\ln N(\text{parent})}{N(n)}}
\end{equation}
where $V(n)$ is cumulative reward, $N(n)$ is visit count, $N(\text{parent})$ is the parent's visit count, and $c$ is the exploration constant. Rewards come from the evaluation pipeline and are backpropagated through the tree after each rollout.

\textbf{Code Refinement Agent:}
Generated variants may fail to compile or pass tests. The refinement agent collects compiler and test diagnostics, asks the LLM to repair the candidate, and re-runs validation before the candidate can enter the optimization pool. CodeEvolve supports both a lightweight reflection loop and an MCTS-based repair loop optimized for unit-test pass rate.

\begin{figure}
    \centering
    \includegraphics[width=\linewidth]{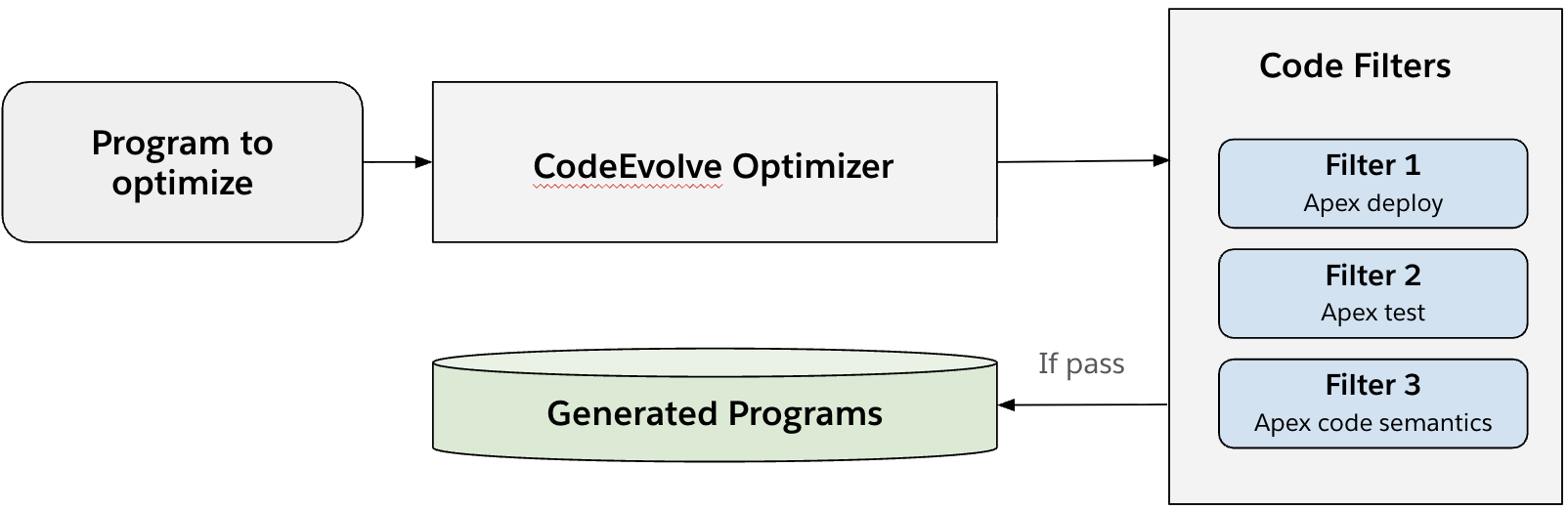}
    \caption{Code filtering pipeline for Salesforce Apex optimization.}
    \vspace{-0.1cm}
    \label{fig:2}
\end{figure}

\textbf{Code Filters:}
Code filters determine whether a generated program is eligible for inclusion in the optimization database. Figure~\ref{fig:2} illustrates the Salesforce Apex filtering pipeline.

Filters include build verification, unit tests, static checks, and optional LLM-based semantic judges. Candidates below the configured thresholds are discarded.

\subsection{Evaluation Pipeline}

CodeEvolve uses language-specific cascades for Java and Apex. Java candidates are evaluated with build validation, unit tests, performance tests, static checks, and LLM-based review. Apex candidates use syntax validation, sandbox tests, governor-limit checks, and Apex-specific review. Candidates must pass early correctness checks before later quality and performance stages are considered. In our experiments, the stage thresholds are $\tau_1 = 0.5$, $\tau_2 = 0.75$, and $\tau_3 = 0.9$, with LLM-based evaluator scores weighted by $\alpha = 0.1$.

\subsection{Algorithm Details}

The main evolutionary loop repeatedly selects a parent, builds a context-aware prompt, asks an LLM to propose a modification, evaluates the child, and inserts only valid candidates into the population. Population diversity is maintained with island-based sampling and periodic migration. The MCTS variant replaces parent sampling with tree-based selection and backpropagates evaluation rewards after each rollout. Appendix~\ref{app:algorithms} gives the full pseudocode and hyperparameters.

\subsection{Code Block Labeling and Targeted Optimization}

CodeEvolve supports targeted optimization through \texttt{EVOLVE-BLOCK} markers, which identify the code region that the optimizer is allowed to modify while preserving surrounding code as read-only context. This interface lets developers or profiling-based target selection constrain the search to a method or small code block rather than an entire source file. Appendix~\ref{app:targeted-interface} provides the marker format and practical usage guidelines.

\section{Ablation Study}

To evaluate the contribution of each key component in CodeEvolve, we conduct an ablation study on Salesforce Apex code optimization, comparing four algorithm configurations over 20 iterations. All experiments target the automated improvement of Apex programs against a set of unit tests and KPI metrics. Results are reported as mean $\pm$ standard deviation across multiple runs.

\begin{enumerate}
    \item \textbf{Original}: The baseline OpenEvolve implementation without modifications.
    \item \textbf{Original + Valid}: OpenEvolve with a code filter that retains only programs passing all unit tests.
    \item \textbf{Improved Context \& Sampling}: The improved implementation with enriched code context and enhanced population sampling strategies.
    \item \textbf{Final (Improved Context \& Sampling + MCTS)}: The full CodeEvolve system, adding the MCTS-based code refinement agent on top of the improved context and sampling.
\end{enumerate}

The KPI score is the \texttt{combined\_score} metric produced by the cascaded evaluation pipeline (Section~3.3). This score aggregates normalized outputs from each evaluation stage. When the combined score is unavailable, the fitness falls back to the arithmetic mean of normalized component scores including code quality, CPU efficiency, memory efficiency, scalability, complexity, and best practices compliance, each normalized to $[0, 1]$.

We evaluate three metrics: (1) the number of generated Apex programs passing all unit tests, reflecting the fraction of valid candidates produced; (2) the average KPI score across all generated programs, measuring overall optimization quality; and (3) the best KPI score achieved, capturing peak optimization performance.

\begin{table}[ht]
\centering
\caption{Ablation study results after 20 iterations (mean $\pm$ std).}
\label{tab:ablation}
\resizebox{\textwidth}{!}{%
\begin{tabular}{lccc}
\toprule
\textbf{Algorithm} & \textbf{\# Programs Passing All Tests} & \textbf{Average KPI Score} & \textbf{Best KPI Score} \\
\midrule
Original                                  & $7.875 \pm 4.155$ & $0.6505 \pm 0.0854$ & $0.9415 \pm 0.0097$ \\
Original + Valid                          & $9.750 \pm 3.496$ & $0.8521 \pm 0.0407$ & $0.9419 \pm 0.0069$ \\
Improved Context \& Sampling              & $13.62 \pm 2.722$ & $0.8725 \pm 0.0417$ & $0.9470 \pm 0.0057$ \\
Final (Improved Context \& Sampling + MCTS) & $19.50 \pm 0.926$ & $0.8977 \pm 0.0251$ & $0.9495 \pm 0.0034$ \\
\bottomrule
\end{tabular}%
}
\end{table}

The results in Table~\ref{tab:ablation} demonstrate the additive benefit of each component. Filtering to only valid Apex programs (Original + Valid) substantially improves the average KPI score from $0.6505$ to $0.8521$, confirming that retaining invalid candidates harms population quality. Adding improved context and sampling further increases both the number of valid programs and the average KPI score. The full system with MCTS achieves the highest performance across all three metrics, reaching an average KPI of $0.8977$ and nearly eliminating variance in the number of valid programs ($19.50 \pm 0.926$ out of 20 iterations), demonstrating the effectiveness of the MCTS-based code refinement agent for consistent, high-quality optimization of Apex code.

\section{Experiments on Real-World Applications}

\subsection{Experimental Setup}

We evaluate CodeEvolve on performance-critical Java code from the Salesforce Monolith, a large-scale enterprise Java codebase used in production. Candidate functions are identified using CodeEvolve's runtime profiling enrichment module (Section~3.1), which profiles production execution traces via JFR, constructs weighted component graphs reflecting execution frequency and cumulative cost, and applies dual-criteria pruning to surface methods with the highest optimization potential. From the full codebase, the profiling enrichment module identified 7 representative hotspot functions spanning diverse optimization scenarios: collection filtering, string mutation, hash computation, object construction, equality checking, and record serialization. These functions were selected because they exhibited both high execution frequency and significant per-invocation cost in the weighted component graph, making them high-value targets for performance improvement.

For each function, we compare two LLM-based optimization approaches:

\begin{itemize}
    \item \textbf{Runtime Profiling Bottleneck Detector (RPBD):} The LLM-based optimization component integrated into CodeEvolve's runtime profiling enrichment module. Given a target function and its surrounding context, RPBD generates a single optimized variant in one LLM call, focusing on micro-optimizations such as eliminating unnecessary method calls, replacing expensive API patterns, and reducing object allocations. RPBD is evaluated using Claude~4~Sonnet.
    \item \textbf{CodeEvolve:} Our full CodeEvolve framework configured for Java code optimization, employing iterative evolutionary search with a full evaluation pipeline comprising build validation, unit testing, static complexity analysis, and LLM-based code review to guide multi-generation improvement. CodeEvolve is evaluated across 4 LLMs: GPT-4.1, Llama~4, Mistral, and an internal open-source model.
\end{itemize}

Performance is quantified as the speedup factor, defined as the ratio of baseline to optimized mean execution time, measured by performance tests using real-world production traffic replayed in a sandbox environment. We report the average speedup across all tested LLMs for each method.

\subsection{Results}

Table~\ref{tab:coreapp} reports the average speedup for each method. CodeEvolve outperforms RPBD on 5 out of 7 functions with an overall average speedup of 15.22$\times$, compared to 7.12$\times$ for RPBD. The advantage is most pronounced on \texttt{HBaseConnectionRegistryUtil.filterMap} (90.34$\times$ vs.\ 42.31$\times$) and \texttt{getInstanceIdMutations} (8.44$\times$ vs.\ 1.58$\times$), where iterative evolutionary search identifies non-trivial algorithmic improvements that single-pass generation rarely achieves consistently. CodeEvolve also leads on \texttt{DomainSuffixPair.hashCode} (2.06$\times$ vs.\ 1.32$\times$), the \texttt{ApiName} constructor (1.85$\times$ vs.\ 0.85$\times$), and \texttt{ComposedApiName.equals} (1.39$\times$ vs.\ 1.16$\times$).

RPBD achieves a slightly higher average on two functions where the optimal optimization reduces to a well-known pattern substitution: \texttt{BatchImporter.serializeRowInfo} (1.26$\times$ vs.\ 1.21$\times$) and \texttt{FieldIdentifier.hashCode} (1.38$\times$ vs.\ 1.27$\times$). The margins are small, and both methods converge on the same core techniques in these cases.

\begin{table}[ht]
\centering
\caption{Average speedup across all tested LLMs on Salesforce Monolith hotspot functions. Values below 1.0$\times$ indicate average regression. Speedup is measured via performance tests with real-world production traffic replayed in a sandbox environment.}
\label{tab:coreapp}
\resizebox{\textwidth}{!}{%
\begin{tabular}{lrrl}
\toprule
\textbf{Hotspot Function} & \textbf{Avg Speedup (RPBD)} & \textbf{Avg Speedup (CodeEvolve)} & \textbf{Key Optimization} \\
\midrule
HBaseConnReg.filterMap           & 42.31$\times$ & \textbf{90.34$\times$} & HashSet conversion (O($nm$)$\to$O($n$+$m$)) \\
DynConfig.getInstanceIdMutations & 1.58$\times$  & \textbf{8.44$\times$}  & Two-pass algorithm; eliminate regex loop \\
DomainSuffixPair.hashCode        & 1.32$\times$  & \textbf{2.06$\times$}  & \texttt{enum.ordinal()} + manual hash computation \\
ApiName.\textit{<init>}          & 0.85$\times$  & \textbf{1.85$\times$}  & \texttt{indexOf}+\texttt{substring} vs.\ \texttt{split} \\
ComposedApiName.equals           & 1.16$\times$  & \textbf{1.39$\times$}  & Reference equality short-circuit \\
BatchImporter.serializeRowInfo   & \textbf{1.26$\times$} & 1.21$\times$ & \texttt{replace()} vs.\ \texttt{replaceAll()} \\
FieldIdentifier.hashCode         & \textbf{1.38$\times$} & 1.27$\times$ & Manual hash; eliminate \texttt{Objects.hash()} \\
\midrule
\textbf{Win Rate / Avg. Speedup} & 2/7 $\cdot$ 7.12$\times$ & \textbf{5/7 $\cdot$ 15.22$\times$} & \\
\bottomrule
\end{tabular}%
}
\end{table}

\subsection{Observation}

The results reveal a consistent advantage for CodeEvolve's iterative evolutionary approach, particularly on functions that admit non-trivial algorithmic improvements.

\textbf{Iterative search discovers superior algorithmic strategies.}
The largest performance gaps arise on functions with complex inefficiencies. In \texttt{filterMap}, the stock implementation calls \texttt{List.contains()} inside a loop, incurring O($n \cdot m$) complexity. RPBD averages 42.31$\times$ across models by discovering partially better implementations, but CodeEvolve's iterative search consistently converges on converting the smaller collection to a \texttt{HashSet} and iterating the larger, reducing complexity to O($n + m$) and achieving an average of 90.34$\times$. In \texttt{getInstanceIdMutations}, RPBD averages only 1.58$\times$ because single-pass generation does not reliably replace the regex-based loop. CodeEvolve's evaluation-guided search discovers a two-pass solution that counts zeros first and generates all mutations without regex, achieving an average of 8.44$\times$. These improvements require multi-step reasoning and iterative validation that single-pass generation cannot consistently provide.

This improvement arises because evolutionary search enables \emph{incremental algorithmic discovery}: rather than requiring the LLM to identify the optimal algorithm in a single reasoning step, the population-based approach allows partial improvements to accumulate across generations. The evaluation pipeline provides a fitness gradient that guides the population toward increasingly efficient solutions, with each generation building on validated improvements from the previous one.

\textbf{Evaluation-driven selection improves consistency.}
CodeEvolve's evaluation pipeline automatically discards candidates that fail to compile or regress on unit tests, retaining only valid and improving variants across generations. This mechanism is particularly beneficial on \texttt{ApiName.\textit{<init>}}, where RPBD averages below baseline (0.85$\times$), while CodeEvolve achieves a consistent 1.85$\times$ average by filtering out regressions and guiding search toward verified improvements.

The cascade evaluation pipeline (Section~3.3) plays a dual role: it acts not only as a filter but as a \emph{selection pressure} that creates a gradient toward better solutions. By discarding candidates that regress on any evaluation dimension, the pipeline ensures that the population monotonically improves, a property that single-pass generation fundamentally lacks.

\textbf{Single-pass generation is competitive for simple pattern substitutions.}
On methods \texttt{BatchImporter.serializeRowInfo} and \texttt{FieldIdentifier.hashCode}, both methods converge on the same simple optimizations. The first replaces \texttt{replaceAll()} with \texttt{replace()}, and the second eliminates the array allocation introduced by \texttt{Objects.hash()}. RPBD achieves marginally higher averages on these two functions (1.26$\times$ vs.\ 1.21$\times$, and 1.38$\times$ vs.\ 1.27$\times$), suggesting that when the optimal transformation is straightforward, single-pass generation is sufficient and the overhead of evolutionary search provides diminishing returns.

These two functions exhibit what we term a \emph{flat optimization landscape}: there is essentially a single well-known transformation that achieves near-optimal performance. In such cases, the first valid candidate found by any approach is likely to be optimal, and evolutionary search adds computational overhead without discovering additional improvement pathways. This suggests that a hybrid strategy could further improve cost efficiency by applying single-pass generation to pattern-matchable optimizations and evolutionary search to algorithmically complex targets.

Taken together, these results demonstrate that CodeEvolve's multi-generation approach delivers substantially higher average speedups (15.22$\times$ vs.\ 7.12$\times$) and wins on 5 out of 7 functions, with the advantage widening where algorithmic insight and reliable correctness validation are required.

\section{Discussion}

\subsection{Practical Implications}

The experiments suggest that LLM-based optimization is most useful when it is scoped, validated, and iterative. Runtime profiling reduces wasted search by identifying code that matters for observed workloads. The evaluation pipeline then turns build results, unit tests, static checks, and performance measurements into a selection signal, preventing invalid candidates from shaping later generations. This is particularly important in enterprise settings, where a locally plausible edit can fail because of platform constraints, hidden dependencies, or coding standards.

The results also indicate that not all optimization targets require the same amount of search. Simple pattern substitutions, such as replacing regex-based string operations or avoiding unnecessary allocation, can often be found by a single LLM call. More complex algorithmic changes benefit from iterative search because partial improvements can be preserved, recombined, and refined across generations. A practical deployment can use this distinction to reserve evolutionary search for targets where single-pass generation is unlikely to be sufficient.

\subsection{Limitations and Research Directions}

CodeEvolve currently depends on the quality of its evaluators. Static analysis and LLM-based review are useful but incomplete proxies for semantic equivalence, maintainability, and long-term code health. Stronger semantic checks and workload-aware dynamic profiling would improve confidence in accepted edits.

Scaling to very large codebases also remains challenging. Runtime profiling identifies promising targets, but dependency analysis and context construction become harder when behavior spans many services or modules. Future work should explore hierarchical target selection, richer context summarization, and human-in-the-loop review so that automated optimization better aligns with developer preferences and organizational standards.

\section{Conclusion}

CodeEvolve shows that LLM-based code optimization is more reliable when generation is embedded in a measured search-and-validation loop. Runtime profiling identifies where optimization is likely to matter, evolutionary and MCTS-based search explore candidate edits, and the cascaded evaluation pipeline filters out changes that fail to preserve correctness. This combination allows the system to improve Java and Apex programs while respecting language-specific constraints.

Our evaluation supports three main findings. First, profiling-guided target selection focuses optimization on high-impact code paths without requiring manual bottleneck analysis. Second, validation-first search improves robustness by preventing invalid programs from shaping later generations. Third, iterative search is most valuable when the required optimization is algorithmic rather than a simple pattern substitution. On the Salesforce Monolith, CodeEvolve achieves an average speedup of 15.22$\times$ across seven hotspot functions and outperforms single-pass optimization on five targets.

Future work will focus on richer semantic analysis, better scaling for very large codebases, domain-specific constraints, and human-in-the-loop mechanisms that align automated optimization with developer preferences and organizational coding standards.

\appendix

\section*{Appendix}

\section{Targeted Optimization Interface}
\label{app:targeted-interface}

CodeEvolve supports targeted optimization through code block labeling using \texttt{EVOLVE-BLOCK} markers, consistent with the OpenEvolve implementation. This feature enables developers to specify precise regions of code as the primary focus of evolutionary optimization.

\begin{verbatim}
// EVOLVE-BLOCK-START
// Code that should be optimized goes here
public void targetMethod() {
    // This method will be the focus of optimization
    for (int i = 0; i < items.size(); i++) {
        processItem(items.get(i));
    }
}
// EVOLVE-BLOCK-END
\end{verbatim}

CodeEvolve extends the OpenEvolve approach by supporting optimization at finer granularity, enabling developers to target either a single method or a specific code block rather than the entire source file. Code outside the marked block is preserved to provide necessary context, but it is not modified during the optimization process.

Key guidelines for effective usage include:
\begin{itemize}
\item \textbf{Single target:} Use one \texttt{EVOLVE-BLOCK} per file to maintain optimization focus.
\item \textbf{Method-level granularity:} Prefer wrapping individual methods or small, self-contained code sections.
\item \textbf{Clear boundaries:} Ensure that the marked block contains complete and compilable code units.
\item \textbf{Context preservation:} Code outside the block is preserved for context and remains unchanged during optimization.
\end{itemize}

\section{Representative Prompt Template}
\label{app:prompt-template}

CodeEvolve prompts are generated from a structured template rather than written manually for each target. The template below shows the fields used for Java optimization; Apex prompts use the same structure with Apex-specific constraints and evaluator names.

\begin{verbatim}
You are optimizing one writable code region in a larger codebase.

Goal:
- Improve runtime performance without changing observable behavior.
- Preserve public APIs, exception behavior, and existing tests.
- Return only a minimal patch for the writable region.

Runtime profile:
- Target: <qualified method or code block name>
- Cumulative time: <T(v)>
- Call count: <C(v)>
- Allocation/CPU notes: <optional profiling annotations>

Writable code:
<EVOLVE-BLOCK or method body>

Read-only context:
<imports, signatures, constants, neighboring methods, types>

Evaluation feedback:
<prior test failures, compiler errors, benchmark results, best variants>

Constraints:
- Do not modify read-only context.
- Prefer simple, maintainable changes.
- Avoid introducing new dependencies unless already present.
- Explain the optimization briefly, then provide the patch.
\end{verbatim}

During multi-generation optimization, the \texttt{Evaluation feedback} field is updated with a compact summary of previous edits and scores. Failed candidates contribute diagnostics, while high-scoring candidates contribute short descriptions of the change that improved the objective.

\section{Algorithmic Details}
\label{app:algorithms}

\begin{algorithm}[!ht]
\caption{Weighted Component Graph Construction}
\label{alg:wcg}
\begin{algorithmic}[1]
\REQUIRE Source codebase $S$, profiling log $P$
\ENSURE Weighted Component Graph $G_w = (V, E, W)$
\STATE Parse $S$ to extract the set of code components $V$ (functions, methods)
\STATE Perform static analysis on $S$ to construct the dependency edge set $E$
\STATE Initialize weight map $W \leftarrow \emptyset$
\FOR{each component $v \in V$}
    \STATE Extract cumulative execution time $T(v)$ from $P$
    \STATE Extract call frequency $C(v)$ from $P$
    \STATE $W_v \leftarrow \langle T(v),\; C(v) \rangle$
    \STATE $W \leftarrow W \cup \{W_v\}$
\ENDFOR
\RETURN $G_w = (V, E, W)$
\end{algorithmic}
\end{algorithm}

\begin{algorithm}[!ht]
\caption{Semantic-Aware Context Pruning}
\label{alg:pruning}
\begin{algorithmic}[1]
\REQUIRE WCG $G_w = (V, E, W)$, thresholds $\tau_{\text{time}}$, $\tau_{\text{freq}}$
\ENSURE Context subgraphs $\{G'_t\}$ for each target $t$
\STATE $V_{\text{target}} \leftarrow \{v \in V \mid T(v) \geq \tau_{\text{time}} \;\text{or}\; C(v) \geq \tau_{\text{freq}}\}$
\STATE $\{G'_t\} \leftarrow \emptyset$
\FOR{each target $t \in V_{\text{target}}$}
    \STATE $V_{\text{frozen}} \leftarrow \{n \in V \mid (t, n) \in E \;\text{or}\; (n, t) \in E\}$
    \STATE $V'_t \leftarrow \{t\} \cup V_{\text{frozen}}$
    \STATE $E'_t \leftarrow \{(u, v) \in E \mid u \in V'_t \;\text{and}\; v \in V'_t\}$
    \STATE $W'_t \leftarrow \{W_v \mid v \in V'_t\}$
    \STATE Mark $t$ as \textsc{Active}; mark all $n \in V_{\text{frozen}}$ as \textsc{Frozen}
    \STATE $G'_t \leftarrow (V'_t, E'_t, W'_t)$
    \STATE $\{G'_t\} \leftarrow \{G'_t\} \cup \{G'_t\}$
\ENDFOR
\RETURN $\{G'_t\}$
\end{algorithmic}
\end{algorithm}

\begin{algorithm}[!ht]
\caption{Evolutionary Code Optimization}
\label{alg:1}
\begin{algorithmic}[1]
\STATE Initialize population with the initial program
\FOR{$i = 1$ to $max\_iterations$}
    \STATE Select a parent program using island-based sampling
    \STATE Generate inspirational programs from the archive
    \STATE Construct the LLM prompt using context and optimization objectives
    \STATE Generate a code modification using an LLM ensemble
    \IF{diff-based evolution is enabled}
        \STATE Parse and apply code diffs
    \ELSE
        \STATE Parse the full code rewrite
    \ENDIF
    \STATE Evaluate the child program using a language-specific evaluation pipeline
    \IF{the program fails code filtering}
        \STATE Continue to the next iteration
    \ENDIF
    \STATE Add the child program to the population
    \STATE Update island populations and perform migration if required
    \STATE Save a checkpoint if required
\ENDFOR
\RETURN The best program in the population
\end{algorithmic}
\end{algorithm}

\begin{algorithm}[!ht]
\caption{MCTS Code Optimization}
\label{alg:2}
\begin{algorithmic}[1]
\STATE Initialize the root node with the initial program
\FOR{$i = 1$ to $max\_iterations$}
    \IF{random() < exploitation\_probability}
        \STATE Select a node using an exploitation strategy
    \ELSE
        \STATE Select a node using UCB-based exploration
    \ENDIF
    \IF{node.count $<$ 1}
        \STATE Evaluate the node through rollout
    \ELSE
        \STATE Generate $k$ child nodes using an LLM
        \STATE Select a random child for evaluation
        \STATE Evaluate the selected child through rollout
    \ENDIF
    \STATE Backpropagate the reward through the tree
    \STATE Add the program to the database if valid
\ENDFOR
\RETURN The best program found in the tree
\end{algorithmic}
\end{algorithm}

For the experiments reported in this paper, the evolutionary optimizer uses $K=5$ islands arranged in a ring topology. Programs are assigned to islands via round-robin allocation. Every 50 generations, migration transfers the top 10\% of each island's population to its two adjacent neighbors. Parent selection uses three sampling modes: elite archive sampling with probability 0.7, current-island sampling with probability 0.2, and uniform sampling from all programs otherwise. The MCTS implementation uses $c=1.0$ as the UCB exploration constant.

\bibliography{merged_refs}
\bibliographystyle{plainnat}

\end{document}